# Dependence of the current renormalisation constants on the quark mass

M. Crisafulli,[a]  V. Lubicz,[a]  G. Martinelli[a] [b]  and A. Vladikas[c] *

[a]Dipartimento di Fisica, Università di Roma 'La Sapienza' and INFN, Sezione di Roma,
P.le A. Moro, I-00185 Rome, Italy.

[b]Theory Division, CERN, CH-1211 Geneva 23, Switzerland.

[c]Dipartimento di Fisica, Università di Roma 'Tor Vergata' and INFN, Sezione di Roma II,
Via della Ricerca Scientifica 1, I-00133 Rome, Italy.

We study the behaviour of the vector and axial current renormalisation constants $Z_V$ and $Z_A$ as a function of the quark mass, $m_q$. We show that sizeable $O(am_q)$ and $O(g_0^2 am_q)$ systematic effects are present in the Wilson and Clover cases respectively. We find that the prescription of Kronfeld, Lepage and Mackenzie for correcting these artefacts is not always successful.

The numerical non perturbative estimate of the vector and axial current matrix elements is afflicted by systematic errors due to the finiteness of the lattice spacing $a$. These errors, which are monitored by measuring the current renormalisation constants $Z_V$ and $Z_A$, are of $O(am_q)$ in lattice simulations based on the Wilson action and $O(g_0^2 am_q)$ in those based on the Clover action ($m_q$ is the quark mass). For light quarks, at $\beta = 6.0$, the above effects were found to be $\sim 25\%$ in the Wilson case. The main success of Clover improvement consists in reducing such systematic effects to $\sim 5\%$. However, when we deal with heavy quark masses $m_h$, $O(g_0^2 am_h)$ effects may become relevant even in the Clover case. In this talk we present a preliminary study of the dependence of $Z_V$ and $Z_A$ on the quark mass.

Recently, Lepage, Mackenzie [1] and Kronfeld [2] (abbreviated as KLM) attempted to absorb these artefacts in modified normalisation factors which match the fermion fields to their continuum counterparts. For clarity of presentation, we separate their proposals into two parts:

(1) <u>KLM normalisation</u>: This is the normalisation factor between the free continuum propagator $P_{cont}(t,\vec{p})$ and its Wilson discrete counterpart $P_{latt}(t,\vec{p})$ which are related by $P_{cont}(t,\vec{p}) = 2K(1+am_q)P_{latt}(t,\vec{p})$ at $\vec{p} = \vec{0}$. This suggests that continuum fermion fields are given in terms of lattice ones by

$$\psi_{cont} = a^{-3/2}\sqrt{2K(1+am_q)}\psi_{latt} \qquad (1)$$

where $am_q = 1/(2K) - 1/(2K_{cr})$ and the critical hopping parameter $K_{cr}$ is obtained non perturbatively. We call the above factor the KLM normalisation. The standard normalisation used previously in lattice simulations, $\psi_{cont} = a^{-3/2}\sqrt{2K}\psi_{latt}$, differs from the KLM one by terms of $O(am_q)$.

(2) <u>MFTI normalisation</u>: Mean Field arguments of [1] suggest a further Mean Field Tadpole Improved (MFTI) prescription for relating the lattice and continuum fields:

$$U_\mu \to U_\mu/u_0 \; ; \; K \to \tilde{K} = Ku_0$$
$$am_q \to a\tilde{m}_q = 8K_{cr}[1/(2K) - 1/(2K_{cr})] \qquad (2)$$

where $u_0$ is any reasonable MF estimate of the expectation value of the link (we use $u_0 = 1/(8K_{cr})$ after [1]). This implies the MFTI normalisation

$$\psi_{cont} = a^{-3/2}\sqrt{2\tilde{K}(1+a\tilde{m}_q)}\psi_{latt} \qquad (3)$$

Bernard [3] took up these ideas and applied them to the non perturbative calculation of $Z_V$ from the ratio of the conserved ($V_\mu^C$) to local ($V_\mu^L$) vector current matrix elements. The spatial component of the conserved current $V_k^C$ has the stan-dard KLM normalisation of eq.(1), but the temporal one, $V_0^C$, being point split in time, requires an extra KLM factor [3]. With $P_5 = \bar{\psi}\gamma_5\psi$ denoting the pseudoscalar density, the final KLM predictions for ratios of the 3-point correlation functions are [3]

$$R \equiv \frac{<P_5|\bar{V}_0^C|P_5>}{<P_5|V_0^L|P_5>} = Z_V \frac{2 + am_1 + am_2}{2} \quad (4)$$

which means that an improved estimate of $Z_V$ may be obtained from

$$R_{KLM} \equiv \frac{<P_5|\bar{V}_0^C|P_5>}{<P_5|V_0^L|P_5>} \frac{2}{2 + am_1 + am_2} = Z_V \quad (5)$$

and its MFTI version is given by

$$R_{MFTI} \equiv \frac{<P_5|\bar{V}_0^C|P_5>}{<P_5|V_0^L|P_5>} \frac{2}{2 + a\tilde{m}_1 + a\tilde{m}_2} \quad (6)$$

---

*Talk given by A. Vladikas.

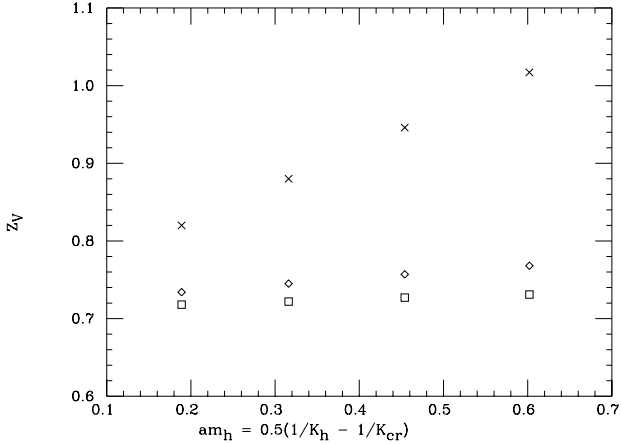

Figure 1. Wilson action estimates of $Z_V$ obtained from the following ratios: $R(\times)$; $R_{KLM}(\diamond)$; $R_{MFTI}(\square)$. The errors are smaller than the symbols.

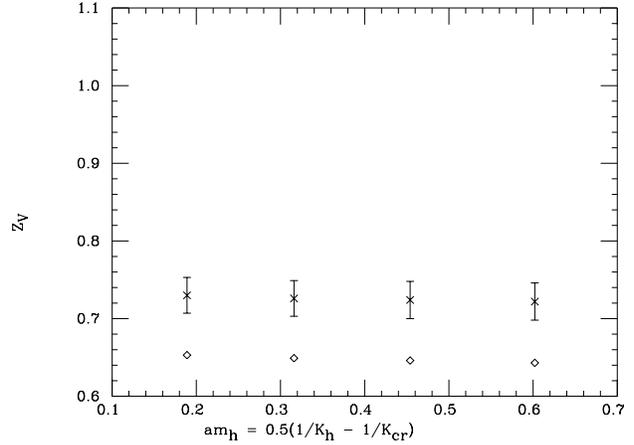

Figure 2. Wilson action estimates of $Z_V$ obtained from the following ratios: $\tilde{R}(\times)$; $\hat{R}(\diamond)$. The errors are only shown when greater than the symbols.

The spatial components, on the other hand, should behave like

$$\tilde{R} \equiv \frac{<P_5|\bar{V}_k^C|P_5>}{<P_5|V_k^L|P_5>} = Z_V \quad (7)$$

and the same is true for the ratio of 2-point correlation functions

$$\hat{R} \equiv \frac{<0|\bar{V}_k^C|\rho>}{<0|V_k^L|\rho>} = Z_V \quad (8)$$

(the above formulae differ from those of [3] because the conserved current used in all our simulations is symmetrised; $\bar{V}_\mu^C(x) \equiv 1/2[V_\mu^C(x) + V_\mu^C(x-\mu)]$). In Figs. 1 and 2 we show the results for the above ratios, as obtained from some ELC data of [4]. The relevant parameters of this simulation were: 20 confs.; $\beta = 6.4$; $V = 24^3 \times 60$. The data shown here are for fixed light quark mass, $K_l = 0.1485$, and varying heavy quark masses $K_h = 0.1275, 0.1325, 0.1375, 0.1425$. The spatial momenta are all set to zero. By comparing the results for $R$, $R_{KLM}$ of Fig. 1 to $\tilde{R}$ of Fig. 2, we see that the KLM normalisation of [1–3] is correcting most of the systematic $O(am_q)$ effects, whereas the $R_{MFTI}$ estimate of Fig. 1 shows that the MFTI correction is of little importance. However, all of these results were obtained from the same matrix elements, $<P_5|V|P_5>$, as opposed to the $Z_V$ estimates of $\hat{R}$ (Fig. 2), obtained from $< 0|V|V >$. The latter estimate of $Z_V$ is incompatible with all the others. This is a typical $O(a)$ effect which the KLM normalisation fails to correct. Recall that, for light quark masses, Clover improvement manages to correct these effects, by implementing a conserved improved current $V_\mu^{CI}$ which differs from $V_\mu^C$ by a total divergence.

We now pass to Clover fermions; here the leading corrections are $O(g_0^2 am_h)$. We will show preliminary results from 40 confs. at $\beta = 6.0$ and $V = 18^3 \times 32$. We have obtained $Z_V$ from ratios of 3-point functions ($R$ of eq.(4)) at zero spatial momenta. The $P_5$ densities are located at $t = 0$ and 16. We have data from correlations obtained at degenerate quark masses $K_h = 0.1150, 0.1200, 0.1250, 0.1330, 0.1425, 0.1432$. We also have results for non-degenerate masses with fixed $K_l = 0.1432$ and $K_h$ varying as above. In this case, $Z_V$ can be checked from the Ward Identity (W.I.) $\nabla_\mu V_\mu^C(x) = \frac{1}{2}[\frac{1}{K_h} - \frac{1}{K_l}]S(x)$ where $S = \bar\psi\psi$ is the scalar density and $\nabla_\mu$ is the asymmetric lattice derivative. Although this W.I. suffers from $O(am_q)$ effects, the $O(am_q)$ improved estimate of $Z_V$ can also be derived from it [5]. Here we only state the final result:

$$Z_V = \frac{1}{8}[\frac{1}{K_h} - \frac{1}{K_l}]$$
$$\frac{< P_5|[2S(x) + S(x+\hat 0) + S(x-\hat 0)]|P_5 >}{\bar\nabla_0 < P_5|V_0^{LI}(x)|P_5 >} \quad (9)$$

The above equation has a symmetric lattice derivative $\bar\nabla_\mu$ and the local improved current $V_\mu^{LI}$; thus it has no $O(am_q)$ terms. Note that all operators (except for $S$) have the standard Clover - rotated fermion fields. The $P_5$'s are at rest. For the axial current renormalisation constant $Z_A$, we have used the estimate obtained from a gauge invariant W.I.[6].

The results are shown in Fig. 3; note that the degenerate mass data does not interpolate smoothly the non degenerate points. This effect is related to the presence of a spectator quark and is currently under investigation [5]. The non zero slopes of the $Z_V$ and $Z_A$ curves of Fig. 3 show the presence of sizeable $O(g^2 am_q)$ linear effects even in the case of the Clover improved matrix elements. A crucial observation from Fig. 3 is that the slopes of the two $Z$'s have opposite signs. Thus, a universal KLM (or MFTI) mass dependent factor cannot flatten out both curves simultaneously. This is another case in which the remedies of refs. [1–3] are shown to be inadequate. More details, as well as more accurate results, consolidating these points, will be shortly presented in [5].

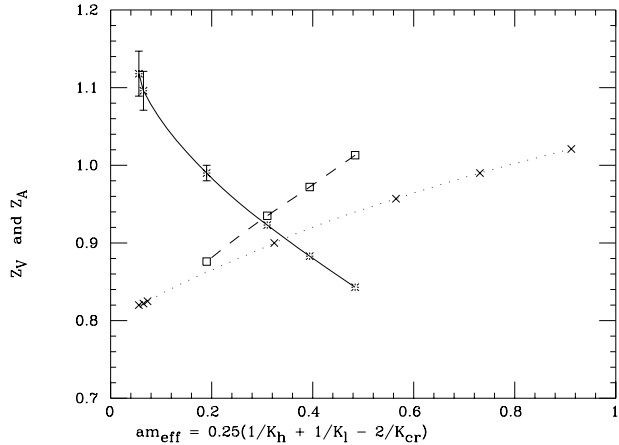

Figure 3. Clover action estimates of $Z_V$ and $Z_A$. $Z_V$ is obtained from $R$ at $K_l = K_h$ ($\times$) and at $K_l \neq K_h$ ($\square$). $Z_A$ is denoted by ($*$). Curves simply guide the eye. The errors are only shown when greater than the symbols.

**REFERENCES**

1. G.P. Lepage and P.B. Mackenzie, Phys. Rev. D48 (1993) 2250, and G.P. Lepage, Nucl.Phys. B(Proc.Suppl.)26 (1992) 45.
2. A.S. Kronfeld, Nucl.Phys. B(Proc.Suppl.)30 (1993) 445.
3. C.W. Bernard, Nucl.Phys. B(Proc.Suppl.)34 (1994) 47.
4. A. Abada et al. Nucl.Phys. B416(1994)675.
5. M. Crisafulli et al., in preparation.
6. G. Martinelli et al. Phys. Lett. B311 (1993) 241.